\documentclass[a4paper, 10pt]{article}
\usepackage[utf8x]{inputenc}
\usepackage{amsmath}
\usepackage{amsfonts}
\usepackage{amssymb}
\usepackage[round]{natbib}
\usepackage{enumerate}
\usepackage{algorithm}
\usepackage{algpseudocode}
\usepackage{graphicx}
\usepackage[colorlinks,bookmarksopen,bookmarksnumbered,citecolor=darkblue, urlcolor=blue, pdfauthor={Tuomas A. Rajala}]{hyperref}

\usepackage[top=3.3cm, left=1.8cm, right=1.8cm, bottom=2.6cm]{geometry}
\usepackage{color}

\newcommand{\x}{\mathbf{x}}
\newcommand{\pp}{\omega}
\newcommand{\y}{\mathbf{y}}

\newcommand{\w}{\theta}
\newcommand{\ov}{\mathbf{o}}

\newcommand{\xii}{{\xi}}
\newcommand{\R}{\mathbb{R}}

\definecolor{darkgreen}{RGB}{0,155,0}
\definecolor{darkblue}{RGB}{0,0,155}
\definecolor{darkred}{RGB}{155,0, 0}

\title{A note on Bayesian logistic regression for spatial exponential family Gibbs point processes}
\author{Tuomas A. Rajala\\ \\Department of Mathematical Sciences\\ Chalmers University of Technology and University of Gothenburg\\ S-41296 Gothenburg, Sweden\\ \texttt{tuomas.rajala@chalmers.se}}

\begin{document}

\maketitle

\begin{abstract}
Recently, a very attractive logistic regression inference method for exponential family Gibbs spatial point processes was introduced. We combined it with the technique of quadratic tangential variational approximation and derived a new Bayesian technique for analysing spatial point patterns. The technique is described in detail, and demonstrated on numerical examples.
\end{abstract}

\section{Introduction}
Gibbs point processes are a popular class of models for interacting events taking place in spatial locations. Parametric inference for these inherently non-independent event models is costly due to analytically intractable normalizing constants in the likelihood. However, due to the recent development of approximate techniques such as the logistic regression technique described by \cite{Baddeley2014a}, the act of modeling is becoming free from the constrains of computational limitations.

In this paper we will describe a Bayesian version of the logistic regression technique of \cite{Baddeley2014a}. Bayesian methodology is not generally shunned upon by point pattern statisticians, so the slow growth in popularity amongst scientists is instead more likely a result of the need to employ MCMC algorithms \citep[see overviews in e.g.][]{handbook2010, moller2007}. MCMC for point process models requires strong expertise in designing the algorithms even for the simplest models, and usually additional, costly simulation loops are needed to evaluate the chains' transitions probabilities. The proposed method requires no simulations and provides posterior distributions in a fraction of the running time of an MCMC algorithm. 

This text is based on the setup described in detail by \cite{Baddeley2014a}, and we often refer to that paper to minimize repetition. Section 2 describes the details of the Bayesian version, and Section 3 depicts some examples on how the Bayesianity can be exploited for a flexible analysis of point pattern data.
\section{Method}
\subsection{Model family and the approximative likelihood}
We assume that we observe a stationary point process in some finite window $W\subset\R^d$. We denote the volume of the window by $V:=|W|$, and write $\pp\subset W$ for the observed, finite configuration of point locations $\pp_i\in W$. The cardinality is denoted by $n=n(\pp):=|\pp\cap W|$. We write $u\in W$ for an arbitrary point in the window. The notation is slightly changed from that of \cite{Baddeley2014a} to simplify the exposition of the model fitting algorithm later on.

The exponential family Gibbs models are defined through a density with respect to a unit rate Poisson process in $\R^d$. In the so called canonical parametrisation, a model of the exponential Gibbs model family has a parametric density of the form
\begin{equation}
\label{eq:f}
\tilde f(\pp|\theta) = \alpha(\theta)^{-1} H(\pp)\exp[\theta^{T}t(\pp)]
\end{equation}
where $\theta\in\R^p$ is a vector of parameters, $t(\pp)=[t_1(\pp)\ldots t_p(\pp)]^T\in\R^p$ is a vector of canonical statistics, function $H(\pp)$ is a baseline or a reference density (mainly to have a hard-core effect), and $\alpha(\theta)$
is the normalizing constant.

The normalizing constant $\alpha(\theta)$ is usually not tractable, hindering direct likelihood based inference. Approximate likelihood inference for Gibbs models is usually based on local dynamics of the process described by the Papangelou conditional intensity, defined at any location $u\in W$ as

\begin{eqnarray}
\label{eq:lambda}
\lambda(u;\pp):=\frac{\tilde f(\pp\cup u)}{\tilde f(\pp\setminus u)}=H(u;\pp)\exp[\theta^T t(u;\pp)])
\end{eqnarray}
with the convention $\pp\setminus u=\pp$ if $u\notin\pp$. Here $H(u;\pp):=1\{H(\pp)>0\}H(\pp\cup u)/H(\pp\setminus u)$ and $t(u;\pp):=t(\pp\cup u)-t(\pp\setminus u)$ are the contributions of the point $u$ to the density.

The logistic likelihood approximation described by \cite{Baddeley2014a} replaces the true likelihood (\ref{eq:f}) with an unbiased estimation equation formally equal to that of standard logistic regression. The method is based on an auxiliary point configuration: First, we simulate a set of dummy locations $d=\{u_j\in W:j=1,...,m\}$ from a known process with a known intensity function $\varrho(u)>0$. Then we concatenate the data and the dummy points to a configuration $\bar\pp=\pp\cup d$, keeping the order, and define a vector of data indicators $y_i:=1\{u_i\in \pp\}$ for each $u_i\in\bar{\pp}$, $i=1,...,n+m$.

Then the likelihood (\ref{eq:f}) is well approximated by a logistic regression density of the form
\begin{equation}
\label{eq:fl}
f(\pp|\theta) = \prod_{i=1}^{n+m}p_i^{y_i}(1-p_i)^{1-y_i}
\end{equation}
where
$$ p_i = \frac{\exp\big[\theta^T t(u_i;\pp)+o_i\big]}{1 + \exp\big[\theta^T t(u_i;\pp)+o_i\big]}$$
and $o_i:=\log[H(u_i;\pp)/\varrho(u_i)]$ are offset terms. Spatial trend components with possible location and covariate dependencies can be included as usual for logistic regression. 

The power of the approach is that the original, computationally hard point process inference problem is approximated by much simpler standard exponential family form which can be solved in the generalized linear model (GLM) framework with standard software. In what follows a variational Bayes method for computing the solution is described.

\subsection{Bayesian inference for logistic regression using variational approximation}
The following variational technique for logistic regression was first described in 1996 and further developed by \cite{Jaakkola2000}, motivated by requirements of fast learning of graphical models. It is based on a tight tangential bound for the function $\log(1+e^{x})$, and with Gaussian priors for $\theta$ the technique leads to tractable Gaussian posteriors. The technique is part of a large group of variational techniques developed for Bayesian data analysis, often grouped under the name Variational Bayes (VB) techniques (see e.g. \cite{Ormerod2010} for an overview).

Write $N:=n+m$. To conform to the usual GLM notation, we will now write $X=[\x_1\ldots \x_N]^T$ for the data matrix with the canonical statistics forming the rows, $\x_i=t(u_i;\pp)$. Using vector notation the logarithm of (\ref{eq:fl}) becomes

\begin{equation}
\label{eq:ll}
\log f(\y|\w) =\sum_{i=1}^{n+m}\left\{y_i (\x_i^T\theta+o_i) - \log[1+\exp(\x_i^T\theta+o_i)]\right\}=\y^T(X\theta + \ov) - 1_N^T\log[1+\exp(X\theta + \ov)]
\end{equation} with $1_N=[1...1]^T$ denoting the $N$ length vector of 1's and $\ov=[o_1 ... o_N]^T$.  The formulation in \cite{Ormerod2010} has now been extended to include the offset terms.

The logarithm terms in (\ref{eq:ll}) make it hard to derive posteriors for $\theta$ as the normalizing integral is not tractable. The method of \cite{Jaakkola2000} works around this by approximating the logarithm terms with tangential lower bounds of quadratic form. The bound is based on the inequality

$$ - \log[1+\exp(\x_i^T\w+o_i)]\ \ \ge\ \ \lambda(\xi_i)(\x_i^T\w + o_i)^2 - \frac{1}{2} (\x_i^T\w+o_i)+\gamma(\xi_i)\quad \forall \xi_i>0$$

where the additional terms are $\lambda(\xi_i)=-\tanh(\xi_i/2)/4\xi_i$ and $\gamma(\xi_i)=\xi_i/2-\log(1+e^\xi_i)+\frac{\xi_i}{4}\tanh(\xi_i/4)$. The $\xi_i$'s are called the variational parameters, and they determine the goodness of the approximation. Equality holds when $\xi_i^2=(\x_i^T\w+o_i)^2$. In vector form, applying the bound to each of the logarithm terms in (\ref{eq:ll}) leads to the quadratic form

\begin{eqnarray*}
-1^T_N\log[1+\exp(X\theta+\ov)]& \ge& (X\theta + \ov)^T\Lambda(\xii)(X\theta + \ov) - \frac{1}{2}1_N^T(X\theta+\ov) + 1_N^T\gamma(\xi)\\
&=&\theta^T X^T\Lambda(\xi)X\theta + 2\ov^T\Lambda(\xi)X\theta + \ov^T\Lambda(\xi)\ov- \frac{1}{2}1_N^T(X\theta+\ov) + 1_N^T\gamma(\xi)
\end{eqnarray*}
where $\Lambda(\xi)=diag(\lambda(\xi_i))$. 

By setting a Gaussian prior $\w\sim N(\mu_0, \Sigma_0)$ the log of the joint density $f(\y,\w)$ in (\ref{eq:ll}) has the closed form lower bound

\begin{eqnarray*}
\log \underline f(\y,\w;\xi) &=& -\frac{1}{2}\w^T[\Sigma_0^{-1}-2X^T\Lambda(\xi)X]\w  \\
&+& [(\y-\frac{1}{2}1_N+2\Lambda(\xi)\ov)^T X + \mu_0^T\Sigma_0^{-1}]\w\\
&+&  1_N^T\gamma(\xi) + \ov^T\Lambda(\xi)\ov +(\y- \frac{1}{2}1_N)^T\ov - \frac{p}{2}\log(2\pi)-\frac{1}{2}\log|\Sigma_0|-\frac{1}{2}\mu_0^T\Sigma_0^{-1}\mu_0 .
\end{eqnarray*}
The bound is proportional to a Gaussian density in $\w$ with posterior covariance matrix and mean given by
\begin{eqnarray*}
\Sigma_\xi^{-1}&=&\Sigma_0^{-1} - 2X^T\Lambda(\xi)X\\
\mu_\xi&=&\Sigma_\xi [(\y- \frac{1}{2}1_N + 2\Lambda(\xi)\ov)^TX+\mu_0^T\Sigma_0^{-1}]^T
\end{eqnarray*}
As these can be computed, we have derived a closed form for the (approximate) posterior distribution.

One detail remains: Both posterior parameters depend on the variational parameters which we need to define. A natural criterion for optimizing the variational parameters is given by the difference between the lower bound of the evidence $f(\y;\xi)$ and the true evidence $f(\y)$, i.e. we aim to minimize the Kullbach-Leibler divergence between the true evidence and the approximation. Due to Gaussianity of $\theta$ the lower bound of the log-evidence can be solved in closed form as
\begin{equation}
\label{eq:evi}
\log \underline f(\y;\xi) = \frac{1}{2}\log|\Sigma_\xi|-\frac{1}{2}\log|\Sigma_0|+1_N^T\gamma(\xi) + \frac{1}{2}\mu_\xi^T\Sigma_\xi^{-1}\mu_\xi - \frac{1}{2}\mu_0^T\Sigma_0^{-1}\mu_0 + \ov^T\Lambda(\xi)\ov +(\y-\frac{1}{2}1_N)^T\ov\ 
\end{equation}
Numerical techniques could now be used for finding $\xi$ that maximizes (\ref{eq:evi}), but \cite{Jaakkola2000} proposed to use a simple expectation-maximization (EM) algorithm instead. With some matrix calculus, especially the linearity of traces, it can be shown that the E-step function has the form
$$Q(\xi'|\xi)=E_{\w|y;\xi} \log \underline f(\y,\w;\xi')=tr\left\{[X(\Sigma_\xi +\mu_\xi\mu_\xi^T)X^T+\ov\ov^T + 2X\mu_\xi \ov^T]\Lambda(\xi')\right\}+1_N^T\gamma(\xi') + \text{const.}$$

As $\lambda(\xi_i)$ is a strictly increasing function of $\xi_i>0$, the optimal values are given by the equation
$$\xi^2 = diag[X(\Sigma_\xi +\mu_\xi\mu_\xi^T)X^T+\ov\ov^T + 2X\mu_\xi \ov^T]\ $$ 
cf. \cite{Ormerod2010} with the current extension of including the offset terms.

\begin{algorithm}
  \caption{Posterior estimation for exponential family Gibbs point processes}
  \label{alg:1}
\begin{algorithmic}[1]
  \State Set $\mu_0, \Sigma_0$, $H$ and $\varrho$
  \State Generate dummy points $d$
  \State Create vector $\y$ and the matrix $X$ 
  \State Initialize $\xi$, $N\times 1$ non-negative vector
  \While{log-evidence is increasing}
    \State Update $\Sigma_\xi^{-1}=\Sigma_0^{-1} - 2X^T\Lambda(\xi)X$
    \State Update $\mu_\xi=\Sigma_\xi [(\y- \frac{1}{2}1_N + 2\Lambda(\xi)\ov)^TX+\mu_0^T\Sigma_0^{-1}]^T$
    \State Update $\xi = \sqrt{diag[X(\Sigma_\xi +\mu_\xi\mu_\xi^T)X^T+\ov\ov^T + 2X\mu_\xi \ov^T]}$
  \EndWhile
  \State Return ($\mu_\xi, \Sigma_\xi$)
\end{algorithmic}
\end{algorithm}

A summary of the technique is given in Algorithm \ref{alg:1}. The computational complexity of the technique differs from that of \cite{Baddeley2014a} only in the logistic regression estimation part so we comment on the complexity of the variational approximation. The variational parameter vector is of size $N=n+m$, and each iteration the inverse and determinant operations are at most of order $O(p^3)$. Since usually $p<<N$ the complexity of the VB algorithm is dominated by the matrix multiplication with complexity of $O(N^2p)$. In comparison, the complexity of solving iterated weighted least squares equation, the default technique in the \texttt{glm}-function used by \texttt{spatstat}-package, is of the same order.

\section{Examples}

\subsection{Strauss process simulations}
To assess the precision of the VB technique we compare it to the default GLM-fitter of the statistical software environment R. The comparison consists of fits to simulations from a range of homogeneous Strauss processes, a model family with repulsive interaction between events' locations. With some fixed parameter $R$ denoting the interaction range, a Strauss process has the canonical statistics in the conditional intensity (\ref{eq:lambda}) of the form
$$t(u;\pp)=[1, n( \pp \cap b(u,R) )]^T$$
where $b(u,R)$ is the ball of radius $R$ centered at $u$. The model has two parameters to estimate, the intensity related parameter $\theta_1$ and the strength of interaction parameter $\theta_2$. Strict definition of the model requires $\theta_2<0$. Note that usually the model is parameterized by $[\beta,\gamma]^T:=\exp(\theta)$. The range parameter $R$ is not of canonical form, so the models are conditional on fixed $R$. 

We simulated 100 realisations of the Strauss model with design $\exp(\theta)\in\{100, 1000\}\times \{.05, .4\}$. To have a realistic range parameter value, we set it to 0.7$R_{max}$, where $R_{max}=2(\frac{2}{\pi^2\lambda})^{1/2}$ is the maximal range under hard packing limit. In setting the range parameters the unknown intensity $\lambda$ was crudely replaced by $\exp(\theta_1)$, leading to $R=\{.06, .02\}$ for small and large $\theta_1$, respectively. The observation window was set to $[0,1]^2$ with additional $R$-dilation for border correction.

We set three types of priors for $\w$:
\begin{enumerate}[a)]
\itemsep 1pt
\item  "flat", $\mu_a=0^T, S_a=diag(10^9)$
\item  "tight" around correct values, $\mu_b=\theta^T$, $S_b=diag(1, .01)$ when $\theta_2=.05$ and $S_b=diag(1, .001)$ when $\theta_2=.4$
\item "tight" around wrong value $\mu_c=\log(2)+\theta^T$ and $S_c=S_b$.
\end{enumerate}
The "tight" wrong priors (c) correspond to stating "expert knowledge" that the true interaction parameter $\gamma=\exp(\theta_2)$ is between .07 and .19 when infact $\gamma=.05$, and between .65 and 1 when $\gamma=.4$. The estimation used the true values of $R$ parameters as is the common practice in similar experiments.

The same dummy point configurations were used for each fitter, and were provided by the current state-of-the art Gibbs model inference function $\texttt{ppm}$ from the \texttt{spatstat}-package \citep{spatstat}.  

\begin{figure}[!ht]
\centering
\includegraphics[scale=0.9]{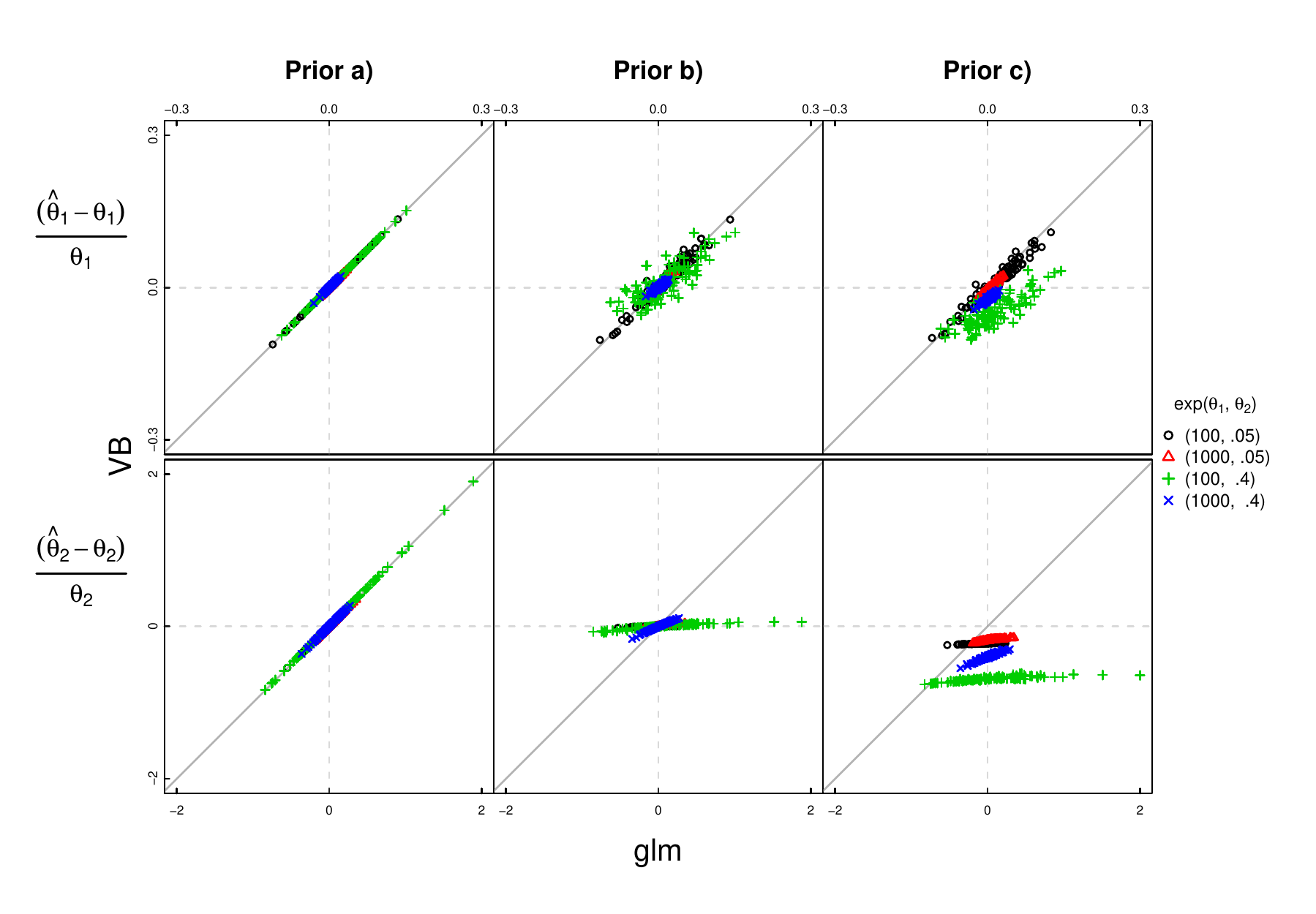}
\caption{Comparison of the VB logistic regression technique and the standard \texttt{glm}-fitter in R. 100 simulations of Strauss model with 4 sets of parameters (see legend). Columns correspond to different priors (see text). Upper and lower row correspond to $\theta_1$ and $\theta_2$, respectively, units are in relative error.}
\label{fig:strauss1}
\end{figure}

Figure \ref{fig:strauss1} shows the estimated parameter values from 100 simulations with the described Bayesian method (VB) plotted against the standard GLM-fitter \texttt{glm} of the R-software as used by the \texttt{ppm}-function. Left column, with the "flat" prior, shows that the VB technique with weak priors induces virtually no extra bias to the estimates, confirming the past experiences of good performance of the tangential approximation \citep{Jaakkola2000}.

The middle column, with tight prior around the true value (b), shows that the for low $<$100 points intensity the estimates are concentrated strongly around the prior (0-line on the figure). Higher intensity diminishes the prior effect. In the right column we see that the wrong prior (c) effect is likewise strong for small intensity and diminishes when more points are available (cf. {\color{darkgreen}$+$} vs {\color{blue}$\times$}). The reduction is higher for less interaction (larger $\theta_2$). It is clear that to fully override strong priors an intensity around $1000$ is not enough, but
for large point patterns (of order $>10^4$) a dubious expert knowledge would most likely be overriden by the data.

\subsection{Mucuous cells with trends}
\begin{figure}[!h]
\centering
\includegraphics[scale=0.6]{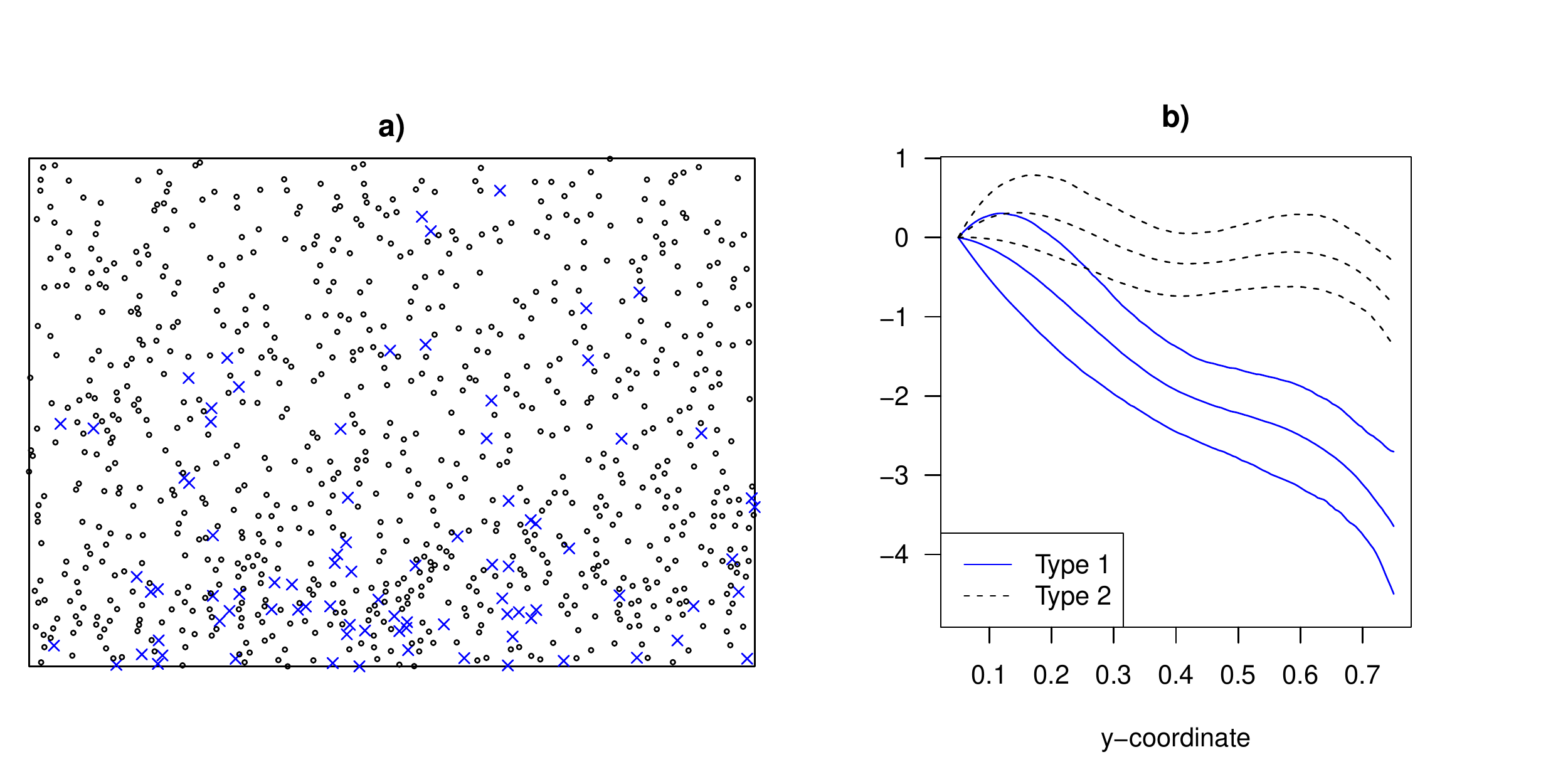}
\caption{a) Mucosa dataset, window $[0, 1] \times [0.05, 0.75]$ type 1 =$\circ$, type 2 = {\color{blue}$\times$}. b) Posterior pointwise 95\%-envelopes for the trends without intercept, as a function of $y$-coordinate.}
\label{fig:mucosa}
\end{figure}

This example revisits the Example 5.4 of \cite{Baddeley2014a}. The data consist of mucous membrane cells of two types, see Figure \ref{fig:mucosa}a). The interesting question is whether the two cell types have a different trend component, indicating that the intensities are not proportional. The trend model for each type is a fourth degree polynomial in the vertical coordinate. A Strauss component is set for cross-type interaction, with range fixed to $R=0.008$. 

Figure \ref{fig:mucosa}b) depicts the posterior pointwise 95\%-envelopes for the trends without intercept term, derived using 1000 simulations from the posterior of the 10-dimensional trend parameter vector. The trends are clearly separate, especially at large $y$-coordinate locations. A confirmation is given by the Bayes factor $26$ against a model with a shared trend.

Figure \ref{fig:mucosa}b) answers the same question as Figure 2b) presented by \cite{Baddeley2014a}, namely what are the pointwise confidence regions for the trends given the data. To derive the regions \cite{Baddeley2014a} deploy a parametric bootstrap scheme where the fitted model is repeatedly simulated. Proper simulation of a point process model is in general not trivial, and can in fact be very costly. In our case we need only simulate a 10-dimension Gaussian random variable, a comparatively trivial execution with very little cost.

\subsection{Inference on an unknown interaction function}
For the so-called pairwise interaction Gibbs model family an often used alternative form of the density is
$$\tilde f(\pp;\theta) \propto \prod_i \phi_0(\pp_i)\prod_{i<j}\phi_\theta(d_{ij}),\quad d_{ij}:=||\pp_i-\pp_j||$$
where  $\phi_0:\R^{d}\rightarrow\R^{+}$ is related to the trend ("first order interaction") and $\phi_\theta:\R^{+}\rightarrow\R^{+}$ is called the (pairwise) interaction function. The connection to exponential form is $\prod\phi_\theta(d_{ij})=\exp[\sum_{i<j}\log\phi_\theta(d_{ij})]$. For example, the Strauss model belongs to this family with the interaction function $\phi_\theta(r)=\exp(\theta_2)^{1(r<R)}$. 

If we do not have a good idea of the shape of the interaction function, i.e. a priori we do not set a specific interaction model, we can approximate it by a step function: Given a set of ranges $0=r_0<r_1<r_2<...<r_K=r_{\max}$, a function
\begin{equation}
\label{eq:psi}
\tilde{\psi_\theta}(r):=\sum_{k=1}^{K}1\{r_{k-1}\le r <r_k\}w_k=\sum_{k=1}^{K}h_k(r)w_k
\end{equation}
with some weights $w_k\in \R$ can be constructed so that it converges to $\log\phi(r)$ as $K\rightarrow \infty$. 

With a Gaussian prior for the weights $\theta_{2:(K+1)}=\bf{w}$, the conditional intensity of the approximating semi-parametric model becomes
$$ \lambda(u_i;\pp) = \exp(\text{1st order term } + {\bf w}^T \psi_i)$$
where $\psi_i=[\sum_j h_1(d_{ij}) ... \sum_j h_K(d_{ij})]^T$ with the coding that we exclude self-interaction and interaction to dummy points, i.e. $h_k(d_{ij})=0$ whenever $i=j$ or $j>n$. The above conditional intensity is that of an exponential model so we are back at the logistic regression situation.

Within the Bayesian framework, we can impose soft constraints on the weights via priors. For example, a typical repulsive interaction function is at $r=0$ equal to 0 and for short ranges close to 0, and at $r_{max}$ interaction converges to 1 (no interaction) as stability conditions require the interaction to be of finite range. We can set the weights to be conditional on these bordering values, and connect the intermediate weights by imposing off-diagonal covariances in Gaussian process fashion, using some suitable kernel function. Note that we are not modeling the interaction function with a multi-scale family but merely approximating it like \cite{Heikkinen1999} did using MCMC. In their implementation the weights were strictly increasing, but unfortunately it is not possible to formulate such hard conditioning in the prior construction.

\begin{figure}[!h]
\centering
\includegraphics[scale=0.6]{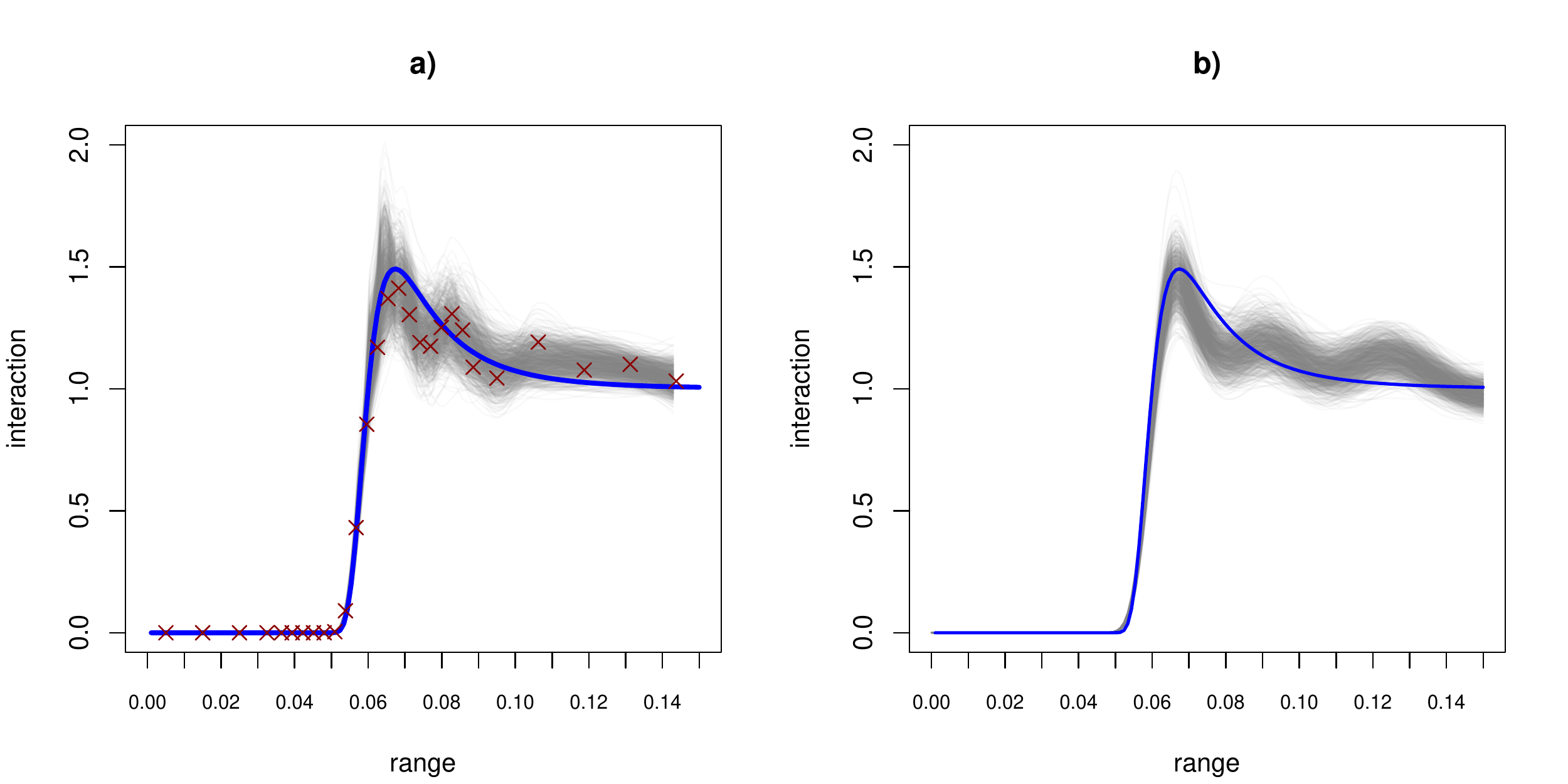}
\caption{a) Interaction function estimation using a step-function with a smoothing prior. True interaction used in simulation ({\color{blue}blue}, thick line), 1000 posterior simulations of the step function, LOESS-smoothed (light gray lines), and the posterior means of the step function weights ({\color{darkred}red $\times$}). b) Interaction function estimated using 50 square exponential basis functions.}
\label{fig:stepfunction}
\end{figure}

In Figure \ref{fig:stepfunction}a) the blue thick line depicts the interaction function of the \cite{LennardJones1924} model, a model motivated by atomic and molecular interactions. A simulation from the model was generated (not shown) in a window $[0,2]^2$ with first order intensity related term $\theta_1=\log(200)$. Overlaid on the true function are LOESS-smoothed samples from the posterior of the step-function (\ref{eq:psi}), along with the posterior mean of $w_k$'s. The estimate of the interaction function is quite acceptable, and an estimate of the characteristic repulsion range with certainty interval can be inferred from the posteriors. Via the formula for the Lennard-Jones interaction function's maxima, the mean estimate of the range is $.058$ with 95\% central confidence region $[.057,.062]$, while the true characteristic range used in the simulation was $.06$.

As a further thought: The step-function approximation can also be extended to the general basis extension approximation. We already did a similar non-parametric approximation for the unknown trend functions in the mucous example, using polynomial basis functions. Figure \ref{fig:stepfunction}b) depicts the estimate for the same Lennard-Jones interaction function using 50 squared-exponential basis functions as $h_k$, using the same $r$-grid for centers as for the step functions. The results are again quite good, considering no particular information of the original potential's shape is used. Characteristic range estimate is also very good, $.059\pm .010$. 

The resolution of the $r$-grid and the basis function width, both corresponding to the level of smoothness of the unknown function is an issue that has been discussed heavily in literature and therefore beyond this text. 


\section{Discussion}
In this exposition we combined two approximation techniques to get a computationally cheap Bayesian method for analysing exponential family Gibbs point processes. The first of the two approximations, and in all accounts the more important one in point pattern context, draws the connection between the conditional intensity of a process and the logistic regression, as described by \cite{Baddeley2014a} and apparently discovered by others during the past decade (personal communications). As it is superior to the most commonly used Poisson approximation to the pseudo-likelihood, the connection is currently the most efficient method for estimating the parameters of these models and is opening interesting doors for point pattern analysis and modeling.

We developed a Bayesian version of the logistic regression method by recalling the second approximation from machine learning literature, a technique that has seen surprisingly little use in the hands of statisticians. The experience on approximating the logistic likelihood using a tangential functions has a good track record in machine learning community, where feasibility, performance and scalability are valued, and the examples shown here only reinforce that optimism.

The shown examples are not particularly new. Bayesian inference for unknown interaction functions and posterior analysis of trend components have been described earlier by e.g. \cite{Heikkinen1999},\cite{Berthelsen2003} and \cite{Bognar2005}. However, most of Baysian analysis of point pattern data are based on MCMC, which is often time consuming for exponential point process models, involving repeated simulations of the process. The described method avoids all such simulations, thus providing real savings on time and computations. And with a full posterior distribution for the parameters at hand some further computational strain can be alleviated in the data analysis over the non-Bayesian counterparts. In the analysis of the mucuous data, for example, the bootstrap simulations of a bivariate Gibbs point process was replaced by considerably easier simulations of a Gaussian vector.

Hopefully this discussion increases the interest of statistical community on the developments happening in other data-oriented fields with similar computational problems. The emphasis on analysis details might be different, but the numerical problems are parallel and we should use that connection to everyones advantage.


\section{Acknowledgements}
The author has been financially supported by the Knut and Alice Wallenberg Foundation.

\bibliographystyle{plainnat}
\bibliography{rajala_gibbs_vb_logistic_031114}

\begin{thebibliography}{10}
\providecommand{\natexlab}[1]{#1}
\providecommand{\url}[1]{\texttt{#1}}
\expandafter\ifx\csname urlstyle\endcsname\relax
  \providecommand{\doi}[1]{doi: #1}\else
  \providecommand{\doi}{doi: \begingroup \urlstyle{rm}\Url}\fi

\bibitem[Baddeley and Turner(2005)]{spatstat}
A.~Baddeley and R.~Turner.
\newblock Spatstat: an {R} package for analyzing spatial point patterns.
\newblock \emph{Journal of Statistical Software}, 12:\penalty0 1--42, 2005.
\newblock URL \url{http://www.jstatsoft.org}.

\bibitem[Baddeley et~al.(2014)Baddeley, Coeurjolly, Rubak, and
  Waagepetersen]{Baddeley2014a}
A.~Baddeley, J.~Coeurjolly, E.~Rubak, and R.~Waagepetersen.
\newblock {Logistic regression for spatial Gibbs point processes}.
\newblock \emph{Biometrika}, 101:\penalty0 377--392, 2014.

\bibitem[Berthelsen and M{\o}ller(2003)]{Berthelsen2003}
K.~Berthelsen and J.~M{\o}ller.
\newblock {Likelihood and Non-parametric Bayesian MCMC Inference for Spatial
  Point Processes Based on Perfect Simulation and Path Sampling}.
\newblock \emph{Scandinavian Journal of Statistics}, 30:\penalty0 549--564,
  2003.

\bibitem[Bognar(2005)]{Bognar2005}
M.~Bognar.
\newblock {Bayesian inference for spatially inhomogeneous pairwise interacting
  point processes}.
\newblock \emph{Computational Statistics \& Data Analysis}, 49:\penalty0 1--18,
  2005.
\newblock ISSN 01679473.

\bibitem[Gelfand et~al.(2010)Gelfand, Diggle, Fuentes, and
  Guttorp]{handbook2010}
A.~Gelfand, P.~Diggle, M.~Fuentes, and P.~Guttorp, editors.
\newblock \emph{Handbook of Spatial Statistics}.
\newblock CRC Press, 2010.

\bibitem[Heikkinen and Penttinen(1999)]{Heikkinen1999}
J.~Heikkinen and A.~Penttinen.
\newblock {Bayesian smoothing in the estimation of the pair potential function
  of Gibbs point processes}.
\newblock \emph{Bernoulli}, 5:\penalty0 1119--1136, 1999.
\newblock URL \url{http://projecteuclid.org/euclid.bj/1143122305}.

\bibitem[Jaakkola and Jordan(2000)]{Jaakkola2000}
T.~Jaakkola and M.~Jordan.
\newblock {Bayesian parameter estimation via variational methods}.
\newblock \emph{Statistics and Computing}, 10:\penalty0 25--37, 2000.
\newblock URL \url{http://www.springerlink.com/index/V651T420415P2312.pdf}.

\bibitem[Lennard-Jones(1924)]{LennardJones1924}
J.~E. Lennard-Jones.
\newblock {On the Determination of Molecular Fields}.
\newblock \emph{Proceedings of the Royal Society A: Mathematical, Physical and
  Engineering Sciences}, 106:\penalty0 463--477, 1924.
\newblock ISSN 1364-5021.

\bibitem[M{\o}ller and Waagepetersen(2007)]{moller2007}
J.~M{\o}ller and R.~Waagepetersen.
\newblock {Modern Statistics for Spatial Point Processes}.
\newblock \emph{Scandinavian Journal of Statistics}, 34:\penalty0 643--684,
  2007.
\newblock ISSN 0303-6898.

\bibitem[Ormerod and Wand(2010)]{Ormerod2010}
J.~Ormerod and M.~Wand.
\newblock {Explaining variational approximations}.
\newblock \emph{The American Statistician}, 64:\penalty0 140--153, 2010.
\newblock URL
  \url{http://amstat.tandfonline.com/doi/abs/10.1198/tast.2010.09058}.

\end{thebibliography}

\end{document}